\documentclass[preprint,12pt]{elsarticle}

\usepackage{amssymb}
\usepackage{graphicx}
\usepackage{threeparttable}
\usepackage{multirow}
\usepackage{makecell}
\usepackage{bbding}
\usepackage{amsmath}
\usepackage{amsfonts}
\usepackage{booktabs}
\usepackage{url}
\journal{Applied Acousitcs}

\begin{document}

\begin{frontmatter}



%\title{UTNet: Unified transformer network 
%leveraging acoustic and textual features
%with Confidence-based Fusion 
%for speech emotion recognition }

\title{
%Individual Prior Estimation for Robust Speech Emotion Recognition
%BSTNet: Band-Split Transformer Network for Noise Robust Speech Emotion Recognition
%Exploring the Impact of Speaker Enrolment on Speech Emotion Recognition
%Revisiting the Correlation of Multi-view Acoustic Features for Speech Emotion Recognition
TRNet: Two-level Refinement Network leveraging Speech Enhancement for Noise Robust Speech Emotion Recognition
}

\author[a,b]{Chengxin Chen\corref{cor1}}
\cortext[cor1]{Corresponding author}
\ead{chenchengxin1120@gmail.com}

\author[a,b]{Pengyuan Zhang}
%\ead{zhangpengyuan@hccl.ioa.ac.cn}
\affiliation[a]{organization={Key Laboratory of Speech Acoustics and Content Understanding, Institute of Acoustics, Chinese Academy of Sciences},
%Department and Organization
%addressline={No. 19, North Fourth Ring West Road, Haidian District}, 
            city={Beijing},
            postcode={100190}, 
            country={China}}

\affiliation[b]{organization={College of Electronic, Electrical and Communication Engineering, University of Chinese Academy of Sciences},
%Department and Organization
%addressline={No. 19, Yuquan Road, Shijingshan District}, 
            city={Beijing},
            postcode={100094}, 
            country={China}}
            
\begin{abstract}
One persistent challenge in Speech Emotion Recognition (SER) is the ubiquitous environmental noise, which frequently results in deteriorating SER performance in practice. In this paper, we introduce a Two-level Refinement Network, dubbed TRNet, to address this challenge. 
Specifically, a pre-trained speech enhancement module is employed for front-end noise reduction and noise level estimation.
Later, we utilize clean speech spectrograms and their corresponding deep representations as reference signals to refine the spectrogram distortion and representation shift of enhanced speech during model training. 
Experimental results validate that the proposed TRNet substantially promotes the robustness of the proposed system in both matched and unmatched noisy environments, without compromising its performance in noise-free environments.
\end{abstract}



\iffalse
\begin{highlights}
%\item We propose a novel noise robust speech emotion recognition algorithm, which effectively integrates a pre-trained speech enhancement module for front-end noise reduction and noise level estimation.
%\item A novel framework which effectively integrates speech enhancement is proposed for speech emotion recognition.
\item A novel speech emotion recognition algorithm is proposed, which effectively integrates speech enhancement for front-end noise reduction and noise level estimation.
\item The low-level feature compensation and high-level representation calibration are designed to jointly promote the noise robustness of the system.
\item The proposed method can achieve superior performance in both noisy and noise-free environments.
\end{highlights}
\fi


\end{frontmatter}


\section{Introduction} \label{sec1}
%The goal of speech emotion recognition (SER) is to categorize human speech into a predetermined set of discrete emotion labels.
%, such as \emph{angry}, \emph{happy}, and \emph{sad}.
Speech emotion recognition (SER) has become a hot topic in the speech processing field due to its wide applications, such as mental health care, personal voice assistants, and user preference analysis~\cite{ser_review}.
%Despite of recent progress, existing works mainly focused on  
%Recent progress in deep learning-based approaches have demonstrated promising performance in SER.
%approaches have significantly promoted the performance of SER.
The goal of SER is to categorize human speech into a predetermined set of discrete emotion labels, and numerous deep learning-based approaches have been proposed to break the upper limit of SER performance~\cite{sota1,sota2,sota3,sota4}.
Most of these works focused on SER in relatively ideal experimental scenarios.
In real-world applications, however, speech signals are frequently contaminated by miscellaneous noises, leading to a prominent drop in SER performance.

To increase the robustness of SER in noisy environments, one strategy involves focusing on feature engineering, exploring the design of feature sets that are insensitive to noise contamination~\cite{schuller2006emotion,georgogiannis2012speech,leem2022not}. For example, Leem~\emph{et al.}~\cite{leem2022not} proposed a feature selection framework that automatically assessed the robustness of each acoustic feature against environmental noise. However, directly transferring these methods to current deep learning-based emotion recognition models presents challenges.
An alternative approach is to investigate from the model's perspective, studying the application of data augmentation techniques to expose the model to target noise during training~\cite{lakomkin2018robustness,tiwari2020multi}. For instance, Lakomkin~\emph{et al.}~\cite{lakomkin2018robustness} introduced random background noise addition and simulated reverberation to contaminate clean training data, while Tiwari~\emph{et al.}~\cite{tiwari2020multi} utilized parametric generative models to generate noisy data. Nevertheless, these methods may not be sufficiently effective on testing data contaminated with unseen noise types.
Recent research has explored methods that integrate speech enhancement (SE) with SER models~\cite{triantafyllopoulos2019towards,zhou2020using,chen2023noise}, aiming to increase the robustness of back-end SER models in noisy environments through noise reduction pre-processing.
Although front-end SE modules can improve SER performance to some extent, the enhanced speech signals often exhibit nonlinear distortions such as artifacts in practical applications~\cite{triantafyllopoulos2019towards}.
%Hence, how to make the most of SE modules for noise robust SER needs further investigation.
Therefore, further research is necessary to optimize the utilization of SE modules for noise robust SER.
%Therefore, further research is necessary to better integrate SE modules into SER systems. 

Motivated by the above observations, this paper proposes TRNet, a Two-level Refinement Network for robust SER in noisy environments.
%real-world scenarios characterized by uncertain noise.
TRNet initially estimates a coefficient with respect to the noise level, enabling low-level feature compensation and high-level representation calibration. 
The low-level feature compensation is designed to approximate target speech spectrograms from pairs of noisy and enhanced spectrograms, while the high-level representation calibration is tailored to align the deep representations extracted from both target and approximated spectrograms.
Experimental results demonstrate that TRNet can effectively couple SE and SER modules, increasing the robustness of the system in both matched and unmatched environments while maintaining SER performance in noise-free environments.

% feature selection
% SE model assistance
% two questions: 1) how to relieve the artifacts phenomenon 2) how to couple SE with backend model

\section{Preliminary} \label{sec2}
% 不用再分小节
Assuming the observed signal $\boldsymbol{x} \in \mathcal{R}^N$ is a mixture of the target speech signal $\boldsymbol{x}_s$ and noise signal $\boldsymbol{x}_n$, \emph{i.e.}, $\boldsymbol{x} = \boldsymbol{x}_s + \boldsymbol{x}_n$, the objective of SE is to recover $\boldsymbol{x}_s$ from $\boldsymbol{x}$.
SE serves as an indispensable front-end module in most speech tasks, such as improving the intelligibility of phone conversations~\cite{reddy2021icassp} and the accuracy of automatic speech recognition~\cite{kinoshita2020improving}.
In recent years, deep learning-based SE has become mainstream, with a plethora of high-performance SE algorithms emerging~\cite{wang2018supervised,zheng2023sixty}.
At a higher signal-to-noise ratio (SNR), the gains from noise reduction may be overwhelmed by losses due to signal distortion.
One strategy to address this issue dynamically adjusts the importance of the front-end SE module based on the estimated SNR~\citep{chen2023noise,koizumi2022snri}. 
\iffalse
Let the enhanced signal be $\boldsymbol{x}_e = \mathrm{SE}(\boldsymbol{x})$, the compensated signal is given by:
\begin{align}
\widetilde{\boldsymbol{x}} = c \cdot \boldsymbol{x} + \left(1-c\right) \cdot \boldsymbol{x}_e
\end{align}
where $c\in \left[ 0,1 \right]$ is a coefficient positively correlated with the SNR.
Under ideal conditions, $c = 1$ corresponds to the clean speech signal, resulting in $\widetilde{\boldsymbol{x}}$ being identical to the original input $\boldsymbol{x}$; as the SNR decreases, the expected noisy signal corresponds to smaller $c$, indicating a higher weight for $\boldsymbol{x}_e$.
\fi
Inspired by the idea of SNR estimation, this paper aims to better couple SE and SER modules from the perspectives of both low-level feature compensation and high-level representation calibration.

\section{Proposed method} \label{sec3}
\begin{figure*}[htb]
  \centering
  \scalebox{1.0}
  {\includegraphics[width=\linewidth]{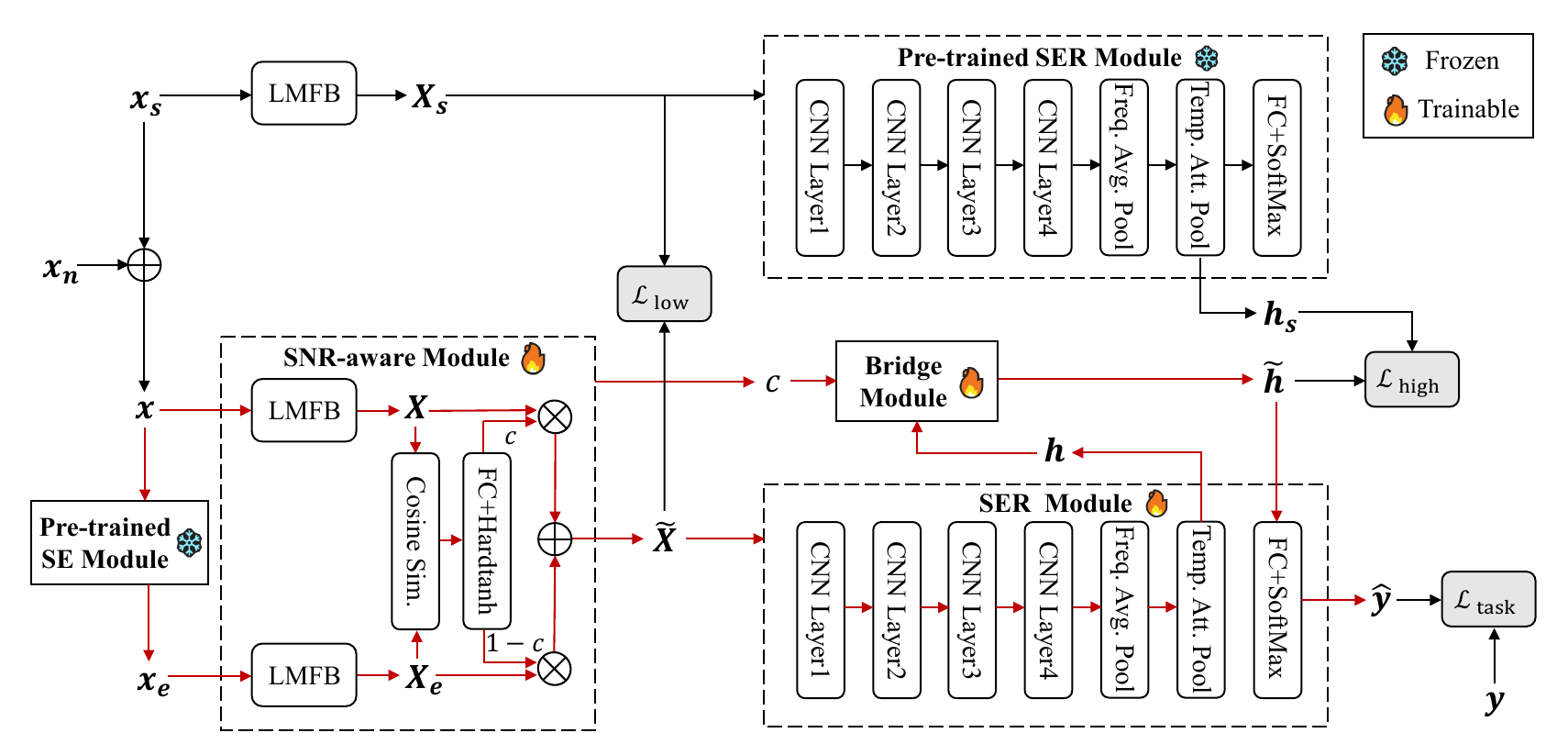}}
  \caption{Diagram of our proposed TRNet. The red arrows are activated during both the training and inference phases, while the black arrows are activated only in the training phase.}
  \label{fig:overview}
\end{figure*}

%front-end feature compensation: SNR score Estimation
%back-end model adaptation: FiLM
As illustrated in Figure~\ref{fig:overview}, the overall workflow of our proposed TRNet comprises five key modules: 
%an SE module and an SNR-aware module are devised for low-level feature compensation, whereas a bridge module and a pair of SER modules are developed for high-level representation calibration.
an SE module, an SNR-aware module, a bridge module and a pair of SER modules.
In the following subsections, we elaborate on the details of each module.

\subsection{SE module}
The SE module employs one of the state-of-the-art architectures, namely the Conformer-based Metric Generative Adversarial Network (CMGAN)~\cite{sherif2024cmgan}. CMGAN was trained on the Voice Bank and DEMAND datasets and the project code of the pre-trained model is publicly available~\footnote{\url{https://github.com/ruizhecao96/CMGAN}}. Previous research has indicated that the joint training of both SE and back-end models often leads to further improved performance in downstream tasks~\cite{li2021espnet}. However, during the training process of TRNet, the parameters of CMGAN are fixed for two reasons: (1) Considering real-world applications, SE serves as a general-purpose front-end module that needs to adapt to multiple downstream tasks. Joint training for the specific SER task may potentially impact the generalization of the SE module for other tasks or unmatched noisy environments; (2) Treating the SE module as a fixed-weight ``black box'' eliminates the need to consider the internal structures of SE models, thus saving computational resources required for training. Note that CMGAN can be replaced by other SE models, such as FSI-Net~\cite{yu2023fsi} or DPT-FSNet~\cite{dang2022dpt}.

\subsection{SNR-aware module}
Given an observed signal $\boldsymbol{x}$ and the corresponding enhanced signal $\boldsymbol{x}_e$, the SNR-aware module aims to dynamically adjust the importance of the SE module.
Firstly, the Log Mel-scale Filter Bank (LMFB) spectrograms $\boldsymbol{X}$ and $\boldsymbol{X}_e$ are computed, where $\boldsymbol{X}, \boldsymbol{X}_e \in \mathcal{R}^{T \times F}$.
Intuitively, the difference between $\boldsymbol{X}$ and $\boldsymbol{X}_e$ increases as the SNR decreases. 
Hence, we calculate the cosine similarity to measure such difference.
Let $\boldsymbol{x}_i$ and $\boldsymbol{x}_i^e$ be the $i$th columns of $\boldsymbol{X}$ and $\boldsymbol{X}_e$, we have
\begin{align}
%\mathcal{D} \left( \boldsymbol{x}_1, \boldsymbol{x}_2 \right) =
d_i = \frac{\boldsymbol{x}_i^\top \boldsymbol{x}_i^e}{\left \| \boldsymbol{x}_i \right \|_2 \left \| \boldsymbol{x}_i^e \right \|_2},
\end{align}
where $d_i$ is the $i$th element of the similarity vector $\boldsymbol{d} \in \mathcal{R}^{F}$, and $\left \| \cdot \right \|_2$ is the $L_2$ norm.
Afterwards, $\boldsymbol{d}$ is passed through a fully connected (FC) layer and a Hardtanh activation function:
\begin{align}
\overline{c} & = \operatorname{FC} \left( \boldsymbol{d} \right), \\
c & = \mathrm{min} \left( \mathrm{max} \left( 0, \overline{c} \right), 1\right),
\end{align}
where $c\in \left[ 0,1 \right]$ is an estimated coefficient positively correlated with the SNR.
Finally, the compensated feature $\widetilde{\boldsymbol{X}}$ is given by
\begin{align}
\widetilde{\boldsymbol{X}} = c \cdot \boldsymbol{X} + \left(1-c\right) \cdot \boldsymbol{X}_e.
\end{align}
At higher SNRs, $c$ is anticipated to approach 1, indicating that $\widetilde{\boldsymbol{X}}$ closely resembles the original input $\boldsymbol{X}$.
As the SNR decreases, the noisy signal corresponds to smaller $c$, resulting in a higher weight for $\boldsymbol{X}_e$.
During model training, the target speech spectrogram $\boldsymbol{X}_s$ is utilized to guide the low-level feature compensation, and the loss function can be calculated as
\begin{align}
\mathcal{L}_\mathrm{low} &= \left \| \boldsymbol{X}_s - \widetilde{\boldsymbol{X}} \right \|^2_2.
\end{align}

\subsection{SER module}
The SER module consists of two major parts. 
Firstly, the acoustic encoder is employed to convert $\widetilde{\boldsymbol{X}}$ into an utterance-level representation, denoted as $\boldsymbol{h}$. Within this encoder, multiple convolutional layers are initially stacked with residual connections to capture local spectrogram characteristics. Subsequently, an average pooling is conducted along the frequency axis, while an attention pooling~\cite{mirsamadi2017automatic} is performed along the time axis to aggregate global information.
Finally, a classifier composed of FC and Softmax layers is employed to project $\boldsymbol{h}$ into a predicted emotion label, denoted as $\hat{y}$. 

As shown in Figure~\ref{fig:overview}, we introduce another identical SER module, which is first pre-trained on the clean emotion dataset and then utilized to extract the utterance-level representation of the target speech, denoted as $\boldsymbol{h}_s$.
During model training, the parameters of the pre-trained SER module are fixed.

\subsection{Bridge module}
While $\widetilde{\boldsymbol{X}}$ is expected to be closer to $\boldsymbol{X}_s$ compared to $\boldsymbol{X}_e$, there still exists a certain level of distortion introduced by the SE module. 
Generally, such distortion becomes more pronounced with increasing SNRs. 
In this section, the bridge module is developed to adjust $\boldsymbol{h}$ according to different SNRs, aiming to directly align the distributions of $\boldsymbol{h}$ and $\boldsymbol{h}_s$ in the emotion space.
%In this way, it can enhance the model's robustness more directly in noisy environments. 
Inspired by Feature-wise Linear Modulation (FiLM) \citep{perez2018film}, we adopt the estimated SNR coefficient $c$ as a constraint to perform an affine transformation on $\boldsymbol{h}$:
\begin{align}
\widetilde{\boldsymbol{h}} = \operatorname{FC}_1 \left( c \right)\cdot\boldsymbol{h} + \operatorname{FC}_2 \left( c \right),
\end{align}
where $\widetilde{\boldsymbol{h}}$ represents the calibrated representation, while $\operatorname{FC}_1$ and $\operatorname{FC}_2$ denote different fully connected layers.
Therefore, the loss function for the high-level representation calibration can be calculated as
\begin{align}
\mathcal{L}_\mathrm{high} &= \left \| \boldsymbol{h}_s - \widetilde{\boldsymbol{h}} \right \|^2_2.
\end{align}

\subsection{Training objectives}
The main task of emotion recognition adopts standard cross-entropy loss function, denoted as $\mathcal{L}_\mathrm{task}$. Eventually, the overall loss function can be computed as
\begin{align}
\mathcal{L} = \mathcal{L}_\mathrm{task} + \alpha \mathcal{L}_\mathrm{low} + \beta \mathcal{L}_\mathrm{high}.
\end{align}
We empirically find that $ \alpha =  \beta = 0.5$ will suffice in our evaluation. 
All the trainable modules of TRNet are jointly optimized by minimizing $\mathcal{L}$.

\section{Experimental setup} \label{sec4}
\subsection{Dataset description}
IEMOCAP~\cite{busso2008iemocap} is a well-benchmarked dataset for SER, which contains 5 sessions and each session is performed by one female and one male actor.
In this research, we considered 4 dominant emotion categories: \emph{Angry}, \emph{Happy}, \emph{Neutral}, and \emph{Sad}.
Following previous works, we merged the utterances labeled \emph{Excited} into the \emph{Happy} category, yielding 5531 speech samples.  

To simulate noisy environments, we adopted another two noise datasets: ESC-50~\cite{piczak2015esc} and MUSAN~\cite{snyder2015musan}. ESC-50 comprises 2000 5-second environmental recordings spanning 50 categories, including animal sounds, natural ambiances, urban noises, \emph{etc}. MUSAN contains approximately 6 hours of noise recordings, encompassing mechanical (\emph{e.g.}, dial tones, fax machine noises) and environmental noises (\emph{e.g.}, thunder, rustling paper). 
We utilized all ESC-50 samples directly, while the noise recordings from MUSAN were segmented into 5-second intervals, resulting in a total of 3860 samples.

\subsection{Implementation details}
The speech samples were resampled at 16 kHz, then the 80-dimensional LMFB features were extracted with 25 ms frame length and 10 ms frame shift.
For parallel computing, the input features of the same minibatch were truncated or padded to a maximum length of 500 frames, and the batch size was set to 32.
For the encoder of SER modules, the filter numbers of the convolutional layers were set as \{32, 64, 128, 256\}, with the kernel sizes of (5, 5) and the strides of (2, 2).
An Adam optimizer with a learning rate of $10^{-3}$ was adopted to optimize the model parameters. The models were trained for a maximum of 80 epochs.

\subsection{Evaluation settings}
In this research, the clean speech dataset is denoted as IEM, while the datasets contaminated by ESC-50 and MUSAN are denoted as IEM-ESC and IEM-MUSAN, respectively. 
To ensure diverse noise exposure across the 5 sessions of IEMOCAP, noise samples were carefully selected to avoid originating from the same source recording. 
The speech samples were contaminated at 5 different SNRs (20 dB, 15 dB, 10 dB, 5 dB, and 0 dB) by random noise samples.
The training data comprised IEM and IEM-ESC, while the testing data encompassed IEM, IEM-ESC, and IEM-MUSAN.

Following previous works, leave-one-speaker-out cross-validation (LOSO CV) was employed in the experiments, where 4 sessions were used for training, while utterances from the remaining two speakers were used for validating and testing, respectively. 
As for performance evaluation, we used the officially recommended metrics, namely unweighted average recall (UAR) and weighted average recall (WAR).

\subsection{Comparative models}
To validate the effectiveness of our proposed method, we implemented three representative models as baselines:
(1) Baseline-c: A basic SER model trained on IEM;
(2) Baseline-n: A basic SER model trained on both IEM and IEM-ESC;
(3) Baseline-e: A basic SER model cascaded with a frozen front-end SE module, trained on both IEM and IEM-ESC;
(4) Baseline-ne: On the basis of Baseline-e, both the noisy and enhanced speech are used as input in a parallel way.
Additionally, two variants of TRNet were configured for comparison:
(1) TRNet w/o $\mathcal{L}_\mathrm{low}$: A model originating from TRNet but excluding the low-level feature compensation;
(2) TRNet w/o $\mathcal{L}_\mathrm{high}$: A model originating from TRNet but excluding the high-level representation calibration.

\section{Results and analysis} \label{sec5}
\subsection{Performance comparison}
First, we perform the quantitative experiments to evaluate the model performance.
According to the results shown in Table~\ref{tab:TRNet_compare}, we conclude the following observations:

(1) In comparison to Baseline-c, both Baseline-n and Baseline-e exhibit a notable enhancement in overall performance 
%ranging from 7\% to 9\% 
across the two noisy environments. However, there is a performance decline 
%of approximately 0.5\% to 1\% 
in the noise-free environment, indicating a trade-off where Baseline-n and Baseline-e prioritize robustness over performance in the noise-free environment. 
%Moreover, Baseline-n surpasses Baseline-e in the matched noisy environment (IEM-ESC), while it falls short of Baseline-e in the unmatched noisy environment (IEM-MUSAN), highlighting a better generalization of Baseline-e.
Additionally, Baseline-ne surpasses both Baseline-n and Baseline-e in the noisy and noise-free environments, but still falls short of TRNet.

(2) The two variants of TRNet perform slightly worse than the original TRNet
%in both noise-free and noisy environments
, yet they consistently outperform Baseline-e. It can be concluded that both low-level feature compensation and high-level representation calibration are beneficial for coupling the SE and SER modules. Moreover, comparing the two variant models reveals that the gain from high-level representation calibration is more pronounced. 
Interestingly, TRNet can achieve even better performance than Baseline-c. This may be attributed to the effective prior knowledge provided by the pre-trained SER module.

\begin{table}[!htbp]
\centering
\caption{Performance comparison of different models. 3 random seeds were used to train each model and the averaged results are reported. IEM-ESC and IEM-MUSAN represent the averaged results at 5 different SNRs. The best results are highlighted in bold.}
\label{tab:TRNet_compare}
\resizebox{\textwidth}{!}{
\begin{tabular}{lcccccc}
%{l p{1.1cm}<{\centering} p{2.3cm}<{\centering} p{2.3cm}<{\centering} p{2.3cm}<{\centering} p{2.3cm}<{\centering}}
\toprule
 \multirow{2}{*}{Methods}  &  \multicolumn{2}{c}{IEM}  & \multicolumn{2}{c}{IEM-ESC} & \multicolumn{2}{c}{IEM-MUSAN} \\
\cmidrule(lr){2-3} \cmidrule(lr){4-5} \cmidrule(lr){6-7}
           &  UAR(\%)   &   WAR(\%)    &  UAR(\%)   &   WAR(\%)  &  UAR(\%)   &   WAR(\%)  \\
\midrule 
Baseline-c    &  56.77   &  54.79   &  45.61  &  43.62  &  44.47   &  42.75  \\
Baseline-n    &  55.85   &  54.48   &  54.22  &  52.65  &  53.39   &  51.84  \\
Baseline-e    &  55.72   &  54.28   &  53.40  &  52.02  &  53.63   &  52.08  \\
%Baseline-ne    &  56.67   &  54.97   &  54.49  &  52.93  &  54.75   &  53.07  
Baseline-ne    &  56.37   &  54.67   &  54.19  &  52.63  &  54.45   &  52.77  \\
\midrule 
TRNet                                   &  \textbf{57.44}   &  \textbf{56.00}   &  \textbf{55.24}  &  \textbf{53.76}  &  \textbf{55.35}   &  \textbf{53.63}   \\   
TRNet w/o $\mathcal{L}_\mathrm{low}$  &  56.70   &  55.02   &  54.77  &  53.13  &  54.90   &  53.02   \\  
TRNet w/o $\mathcal{L}_\mathrm{high}$   &  56.12   &  54.32   &  53.94  &  52.20  &  54.18   &  52.37   \\  
\bottomrule 
\end{tabular}}
\end{table}

\subsection{Computational complexity analysis}
We further compare the computational complexity of different SE-based methods via the metrics of model parameters and multiply–accumulate operations (MACs)~\footnote{\url{https://pypi.org/project/thop/}}. Note that we exclude the computational complexity of the SE module. 
From Table~\ref{tab:complexity}, we find that the MACs of Baseline-ne are twice those of Baseline-e during both the training and inference phases due to the parallel input of noisy and enhanced speech.
By contrast, the extra model parameters and MACs of TRNet are introduced only in the training phase by the pre-trained SER module.

\begin{table}[htbp]
\centering
\caption{Computational complexity of different models.
The model parameters and MACs in the training and inference phases are reported, respectively.}
\label{tab:complexity}
 \scalebox{0.9}{
\begin{tabular}{lcccc}
\toprule
{\multirow{2}{*}{Methods}} & \multicolumn{2}{c}{Model for training}& \multicolumn{2}{c}{Model for inference}\\
\cmidrule(lr){2-3} \cmidrule(lr){4-5} 
                & Params (M)  & MACs (G) & Params (M)  & MACs (G)    \\
\midrule                        
Baseline-e          & 4.60  & 6.32  & 4.60  & 6.32  \\
Baseline-ne         & 4.63  & 12.64  & 4.63  & 12.64  \\
TRNet               & 9.20  & 12.64  & 4.60  & 6.32  \\
\bottomrule
\end{tabular}}
\end{table}

\subsection{Effect of SNR estimation}
\begin{figure*}[htb]
  \centering
  \scalebox{0.8}
  {\includegraphics[width=\linewidth]{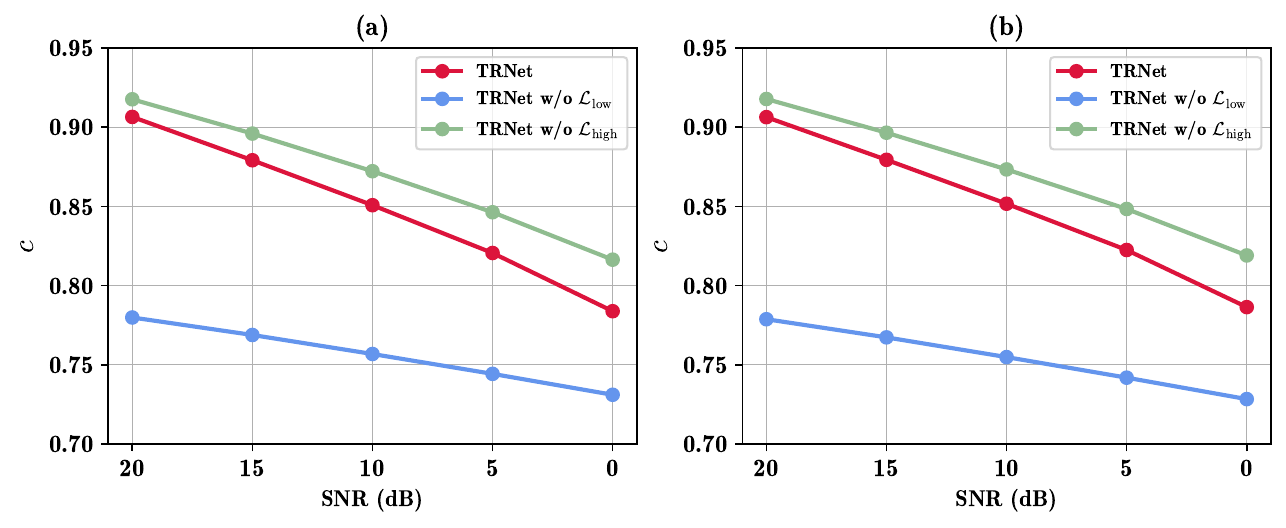}}
  \caption{Estimated SNR coefficients in different noisy environments. (a) IEM-ESC; (b) IEM-MUSAN.}
  \label{fig:TRNet_snr_curve}
\end{figure*}

In this section, we further investigate the role of the estimated $c$ in TRNet and its two variant models. As illustrated in Figure~\ref{fig:TRNet_snr_curve}, we conclude the following observations:
\begin{enumerate}
    \item As expected, $c$ decreases as the SNR decreases, signifying an increase in the weights of enhanced signals. Moreover, similar trends are noted in both matched and unmatched noisy environments, showcasing TRNet's capability to adapt to unseen noise.
    \item 
    %A comparison between the two variant models of TRNet unveils notable discrepancies in $c$ for the low-level feature compensation and high-level representation calibration. 
    %Particularly, the latter one seems to be less sensitive to the variation of $c$.
    In comparison to TRNet w/o $\mathcal{L}_\mathrm{high}$, TRNet w/o $\mathcal{L}_\mathrm{low}$ appears to be less sensitive to $c$ when the SNR varies.
    We speculate the reason is that $c$ functions directly as a linear combination coefficient in the low-level feature compensation, whereas it undergoes an affine transformation in the high-level representation calibration.
\end{enumerate}

\subsection{Visualization analysis}
\begin{figure*}[htb]
  \centering
  \scalebox{1.0}
  {\includegraphics[width=\linewidth]{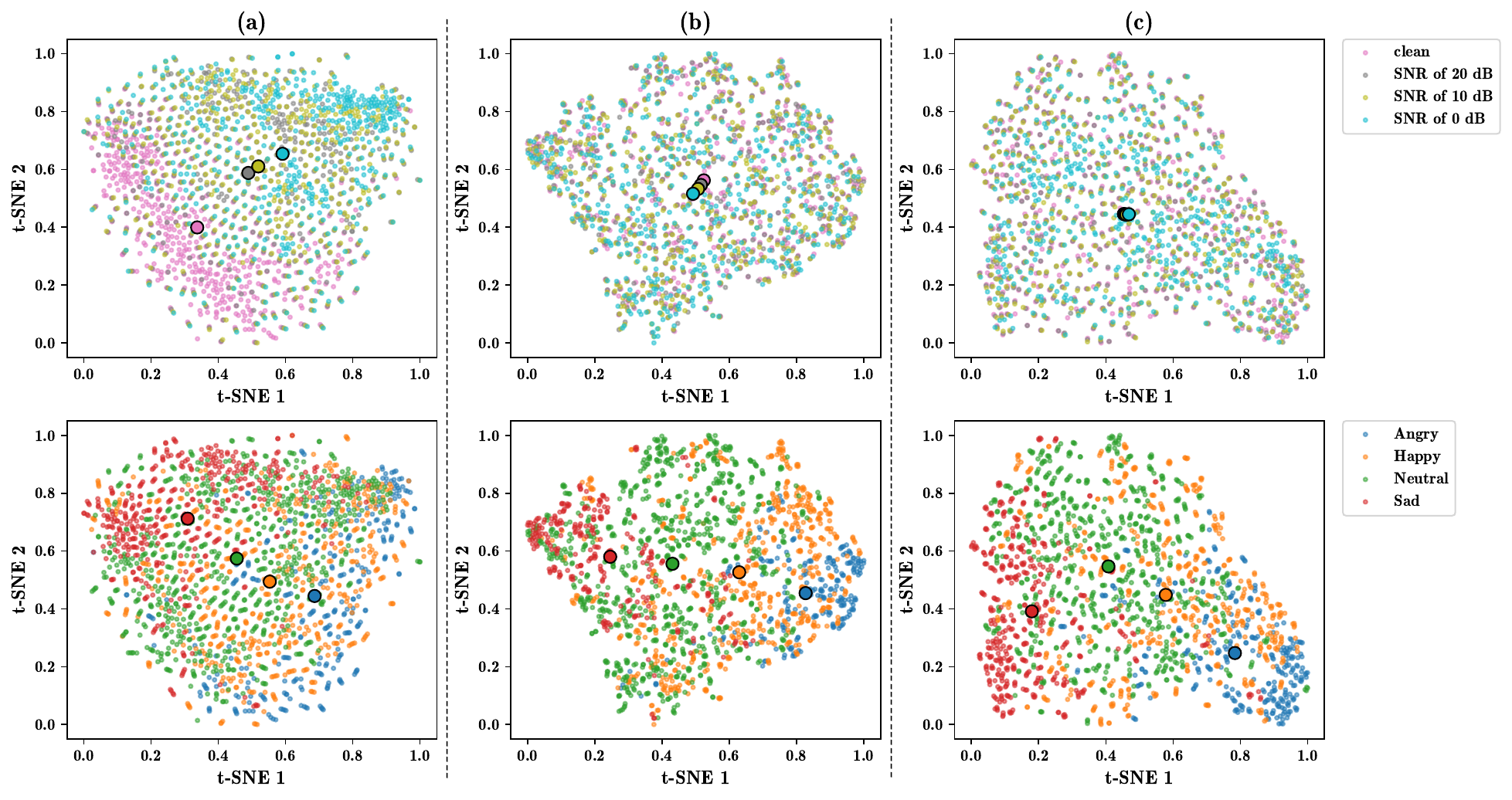}}
  \caption{T-SNE visualization of deep emotion representations. 500 clean utterances were randomly selected and contaminated by noise samples from ESC-50 at different SNRs (20 dB, 10 dB, and 0 dB). The smaller semi-transparent points represent specific samples, while the larger opaque ones correspond to centers of different distributions. (a) Baseline-c; (b) Baseline-e; (c) TRNet.}
  \label{fig:TRNet_ebd_shift}
\end{figure*}

Figure~\ref{fig:TRNet_ebd_shift} visualizes the distribution of deep emotion representations at different SNRs using t-SNE~\cite{van2008visualizing}. 
As shown in the first row, varying SNRs cause significant shifts in the distributional centers for Baseline-c. In contrast, Baseline-e mitigates this phenomenon to a certain extent, and TRNet further aligns the distributional centers at different SNRs. This suggests that TRNet can adeptly adjust the deep emotion representations in various noisy environments and approximate the representations of clean utterances more closely.
In the second row of Figure~\ref{fig:TRNet_ebd_shift}, an evident increase in distance between the centers of diverse emotion categories can be observed for TRNet, which leads to a stronger emotional discrimination within the deep representations.

\section{Conclusion} \label{sec6}
In this paper, we propose TRNet, a noise robust SER algorithm that leverages a pre-trained SE module for front-end noise reduction and noise level estimation. 
Based on the estimated SNR coefficient, the low-level feature compensation and high-level representation calibration are performed, jointly enhancing the robustness of the system against both seen and unseen environmental noise.
Quantitative experimental results demonstrate that TRNet can achieve superior performance in both noisy and noise-free environments.
We further validate the roles of SNR estimation and the characteristics of deep representations through ablation study and visualization analysis.
In future work, we plan to extend the applicability of TRNet to more complex acoustic environments.




\section*{Acknowledgement}
This research was partially supported by the National Key Research and Development Program of China (No. 2021YFC3320103).

\end{document}